\documentclass[pra, twocolumn,amsmath,amssymb,superscriptaddress]{revtex4}
\usepackage{graphicx, color}
\usepackage{dcolumn}
\usepackage{bm}
\usepackage{color}
\begin{document}
\title{The Realization and Detection of Weyl Semimetals and Chiral Anomaly in Cold Atomic Systems } 
\author{Wen-Yu He}
\affiliation{Department of Physics, Hong Kong University of Science and Technology, Clear Water Bay, Hong Kong, China.}
\author{Shizhong Zhang}
\affiliation{Department of Physics and Center of Theoretical and Computational Physic, University of Hong Kong, Hong Kong, China.}
\author{ K. T. Law}\thanks{Correspondence address: phlaw@ust.hk}
\affiliation{Department of Physics, Hong Kong University of Science and Technology, Clear Water Bay, Hong Kong, China.}
\date{\today}

\begin{abstract}
{In this work, we describe a method to realize 3D Weyl semimetal by coupling multilayers of honeycomb optical lattice in the presence of a pair of Raman lasers. The Raman lasers render each isolated honeycomb layer a Chern insulator. With finite interlayer coupling, the bulk gap of the system closes at certain out-of-plane momenta due to Raman assisted tunnelling and result in the Weyl semimetal phase. Using experimentally relevant parameters, we show that both one and two pairs of Weyl points can be realized by tuning the interlayer coupling strength. We suggest that Landau-Zener tunnelling can be used to detect Weyl points and show that the transition probability increases dramatically when Weyl point emerges. The realization of chiral anomaly by using a magnetic field gradient is also discussed.}
\end{abstract}

\maketitle
\section{\bf Introduction}
The search for topological phases that are fully gapped in the bulk and possess topologically protected gapless surface states has been an important subject in physics. Several new topological phases, such as 2D and 3D topological insulators~\cite{Kane,Qi} and quantum Anomalous Hall insulators~\cite{Chang}, have been realized experimentally in condensed matter systems. The gapless surface states in these systems are protected by certain symmetries and the bulk gap~\cite{Kane,Qi}. Nevertheless, the surface states can be removed once the bulk gap in the system is closed. 

Interestingly, it is well known that Weyl semimetals~\cite{Wan,Burkov1,Jiang,Lu,Xu1}, which contain nodal points, the Weyl points, in the energy spectrum also support topologically protected gapless Fermi arcs on the surface. A pair of Weyl points in a Weyl semimetal can be regarded as a pair of topological defects in 3D momentum space with opposite chirality and possess linear energy spectrum near the Weyl points. The Fermi arcs, which connect the Weyl points in the surface Brillouin zone, cannot be removed unless the Weyl points are merged together by strong perturbations~\cite{Wan,Burkov1}. Due to the presence of the Weyl points in the bulk and the gapless Fermi arcs on the surface, Weyl semimetals are expected to exhibit interesting novel phenomena which are absent in fully gapped topological phases such as the chiral anomaly~\cite{Zyuzin,Wang1} and unconventional quantum oscillations~\cite{potter}. Recently the experimental realization of Weyl semimetal phase is under widely studying in real materials~\cite{Hassan, Hong}. 

On the other hand, great advancement has been made in realizing topological phases with cold atoms in optical lattices. Using Raman laser coupling ~\cite{Alba,Goldman1,Beri,Liu1,Aidelsburger1,Kennedy,Liu2} and optical shaking lattice ~\cite{Hauke1,Zheng,Reichl}, Chern insulators have been successfully realized experimentally~\cite{Jotzu,Aidelsburger2} using both schemes. Methods on realizing 3D topological insulators in optical lattices have also been proposed~\cite{Bermudez,Wang2}. In this work, we outline a practical scheme to realize and detect Weyl semimetal phase in optical lattices. 

We base our proposal on the already realized honeycomb optical lattice~\cite{Tarruell,Uehlinger} and show that by introducing non-trivial hopping between the neighboring honeycomb lattice layers, it is possible to generate either one or two pairs of Weyl points in the system. What is important in our condtion is that the momentum transfer $\Delta{\bf p}$ from Raman beams possesses nonzero components both along and perpendicular to the honeycomb lattice planes. Within each individual lattice layer, in-plane momentum transfer generates non-trivial Peierls phases and can drive each layer into a 2D Chern insulator phase. Importantly, when individual layers are coupled with non-trivial phases generated by the Raman beams, the Chern insulators can become trivial for certain out-of-plane momentum $k_z$. Since the bulk gap has to be closed for topological phase transitions from the topologically non-trivial Chern insulators to trivial insulators as a function of out-of-plane momenta $k_z$, there are Weyl points in the momentum space separating the trivial and non-trivial regimes. The number of Weyl points, corresponding to the number of the band-gap closing points as a function of $k_z$, is tunable by changing, for example, the Raman detuning and the interlayer coupling. We suggest that Landau-Zener transitions can be used to detect the emergence of the Weyl points. We also discuss how Landau levels and the chiral anomaly can be realized by introducing a magnetic field gradient in the $z$-direction.

\section{\bf The scheme}
\begin{figure}
\begin{center}
\includegraphics[width=3.3in]{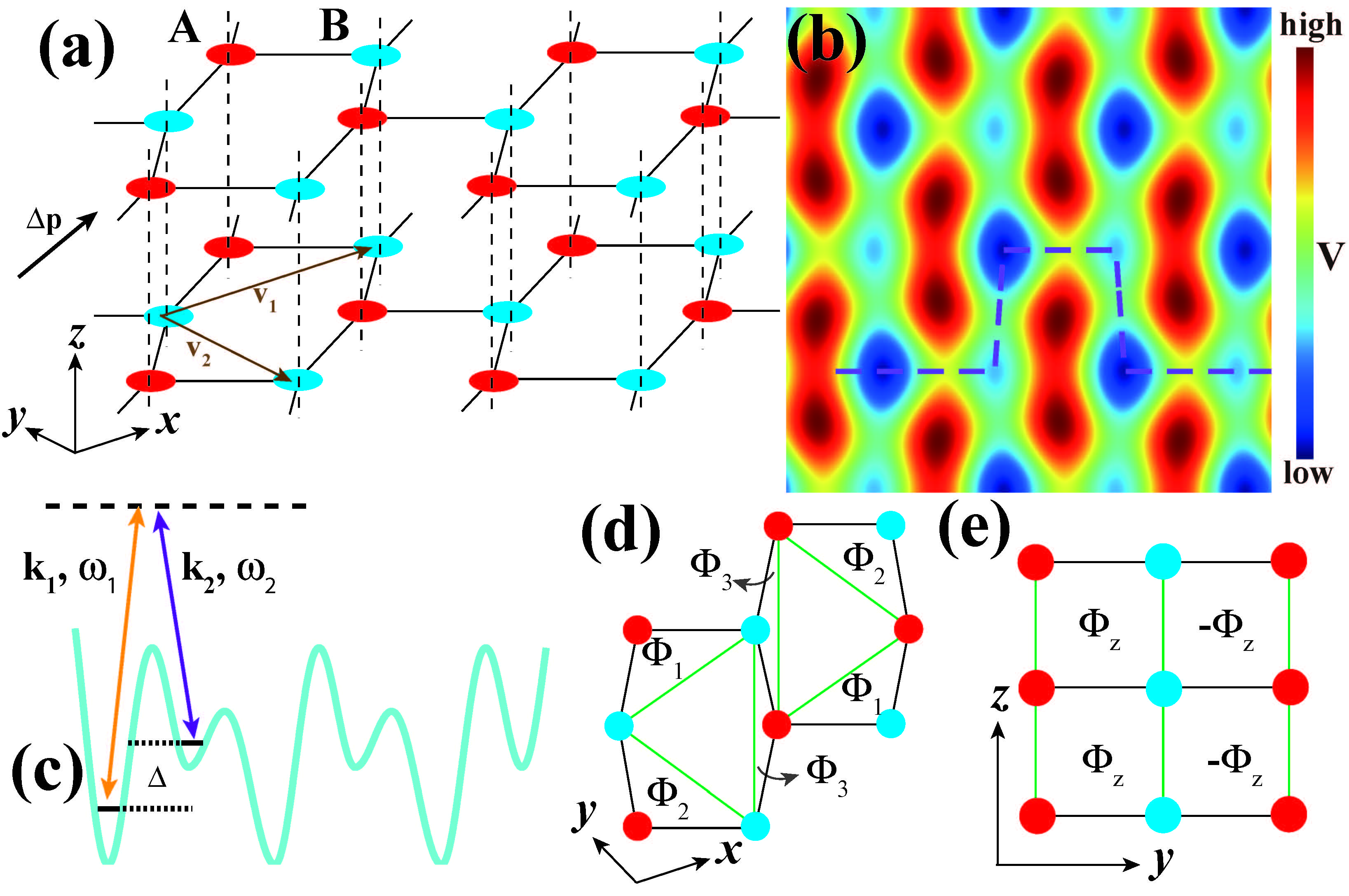}
\caption{(color online) (a) AA-stack of multilayer honeycomb optical lattice. (b) The intralayer optical potential. The bigger dark blue and the smaller light blue pockets represents the A and B sites respectively. (c) The Raman laser assisted resonant tunneling between nearest neighbors. The recoil momentum carried by photons generates Peierls phases in the hopping amplitudes. (d) The synthetic magnetic flux $\Phi_i$ of each triangle resulting from the Peierls phases in the $xy$-plane. (e) The synthetic magnetic flux of each rectangle in the $yz$-plane.}
\label{Fig1}
\end{center}
\end{figure}

We first consider a multilayer honeycomb optical lattice generated by laser beams~\cite{Tarruell,Uehlinger}.
\begin{align}\nonumber
V&=-V_{\bar{X}}\cos^{2}(q\hat{x}+\theta/2 )-V_{X}\cos^{2}(q\hat{x}) -V_{Y}\cos^{2}(q\hat{y})\\
 & -2\alpha \sqrt{V_{X}V_{Y}}\cos(q\hat{x})\cos(q\hat{y}) -V_{\bar{Z}}\cos^{2}(q_{z}\hat{z}),
 \end{align}
where $V_X,V_{\bar{X}}$ and $V_Y$ are intensities of lasers $X,\bar{X}$ and $Y$, pointing along $\hat{x},-\hat{x}$ and $\hat{y}$-directions, respectively. $q$ is the wave vector of the lasers. When \begin{math}V_{\bar{X}}\gg V_{Y} \geqslant V_{Z}\end{math}, a honeycomb lattice is created with two lattice sites $A$ and $B$ in a unit cell, as shown in Fig.\ref{Fig1}a. The energy offset $\Delta$ between A and B sublattices can be tuned through the phase difference $\theta$ between $X$ and $\bar{X}$ lasers. The standing wave $\bar{Z}$, with wave vector $q_z$, stacks the honeycomb optical lattice as seen in Fig.\ref{Fig1}a, and controls the interlayer coupling strength. The energy offset $\Delta$ is tuned to be large enough such that the direct tunnelling between neighboring $A$ and $B$ sites is greatly suppressed in the absence of Raman beams. 

A pair of Raman beams with frequency detuning $\delta\omega\equiv\omega_{1}-\omega_{2}\approx\Delta/\hbar$ are added to restore the resonant tunnelling between neighboring intralayer sites, as illustrated in Fig.\ref{Fig1}c. The momentum transfer $\Delta{\bf p}={\bf k}_1-{\bf k}_2$ associated with the resonant tunnelling introduces Peierls phases $\exp\left[i\Delta{\bf p}\cdot\left({\bf r}+{\bf r}'\right)/2\right]$ between intralayer neighboring sites ${\bf r}$ and ${\bf r}'$. This introduces synthetic magnetic flux into the optical lattices~\cite{Aidelsburger1,Jaksch} as depicted in Fig.\ref{Fig1}d. Using Floquet analysis and a unitary transformation which eliminates the spatial dependence of the Peierls phases (for details, see Appendix~\ref{App-A}), the effective Hamiltonian can be written as
\begin{align}\nonumber
H&=\sum_{ \langle i,j \rangle, l }\left ( t_{ij}^{0}\tilde{b}_{i,l}^{\dagger}\tilde{a}_{j,l}+ { t_{ij}^{a}e^{i\phi _{ij}}\tilde{a}_{i,l}^{\dagger}\tilde{a}_{j,l}} + { t_{ij}^{b}e^{-i\phi _{ij}}\tilde{b}_{i,l}^{\dagger}\tilde{b}_{j,l}}  \right ) \\\nonumber
&+\epsilon \sum_{i,l} \left ({ \tilde{a}_{i,l}^{\dagger}\tilde{a}_{i,l}}\right. \left.{-\tilde{b}_{i,l}^{\dagger}\tilde{b}_{i,l}} \right ) +\sum_{i,l} \left ({ {t_{\perp }^{a}e^{i\phi _{\perp}}\tilde{a}_{i,l}^{\dagger}\tilde{a}_{i,l+1}}}\right.\\
&\left.{+ t_{\perp }^{b}e^{-i\phi _{\perp}}\tilde{b}_{i,l}^{\dagger}\tilde{b}_{i,l+1}}\right )+ h.c..
\end{align}
Here, $i,j$ labels the position of lattice sites within a layer and $l$ is the layer index. $\tilde{a}_{i,l}$ ($\tilde{b}_{i,l}$) annihilates an atom at sublattice A (B) of site $i$ and layer $l$. $t_{ij}^{0}$, which satisfies $t_{i,i-1}^{0}=-t_{i,i+1}^{0}$ due to the Raman fields \cite{Cooper}, denotes the intralayer nearest neighbour hopping amplitude and $t_{ij}^{a}$ ($t_{ij}^{b}$) is the next nearest neighbour hopping amplitude between A (B) sublattice sites. $t_{\perp}^{a}$ and $t_{\perp}^{b}$ are the interlayer hopping amplitudes and $\epsilon=\frac{1}{2}\left(\hbar\delta\omega-\Delta\right)$, originating from the Raman detuning, can be regarded as the on-site energy difference between the sublattices. The Peierls phases $\phi_{ij}$ and $\phi_\perp$ are associated with intralayer and interlayer hopping, generated by the in-plane and the out-of-plane components of $\Delta{\bf p}$, respectively. Explicitly, one finds
\begin{align}
\phi_{ij}=\frac{1}{2}\Delta{\bf p}\cdot({\bf r}_{i,l}-{\bf r}_{j,l}),\phi_{\perp}=\frac{1}{2}\Delta{\bf p}\cdot({\bf r}_{i,l}-{\bf r}_{i,l+1}).
\end{align}
In particular, $\phi_{ij}$ is responsible for making each isolated layer of honeycomb lattices a Chern insulator and $\phi_{\perp}$, which originates from the phase difference between adjacent layers due to nonzero vertical momentum transferred in the Raman process, induces unconventional interlayer coupling which is the key for generating the Weyl semimetal phase. The accumulation of Peierls phases along closed paths gives rise to synthetic magnetic flux patterns as shown in Fig.\ref{Fig1}d and e.

\section{\bf The Weyl points}
To illustrate the emergence of Weyl points, we first consider an ideal version of experimental honeycomb lattice (brickwall lattice). For simplicity, we set $\left |t_{ij}^{0}\right |=t$ and $t_{ij}^{a}=t_{ij}^{b}$ (in the following NNN hopping $t_1$ is along $\bf{v}_1$ or $\bf{v}_2$ while $t_2$ is along $\bf{v}_1-\bf{v}_2$). We use the coordinate system established as shown in Fig.\ref{Fig1}a for convenience and the momentum transfer is set to be $\Delta\mathbf{p}=\left(p_{x},p_{y},p_{z}\right)$. If we further label the $A(B)$ sublattices as corresponding to a pseudo-spin $\boldsymbol\sigma$, then the Hamiltonian takes the following form in momentum space
\begin{align}
H\left ( \mathbf{k} \right )=d_{0}\left ( \mathbf{k} \right )I+\mathbf{d}\left ( \mathbf{k} \right )\cdot\bm{\sigma}   \label{H}
\end{align}
with
\begin{align}\nonumber
d_{0}({\mathbf k})&=2t_{1}\cos k_{x}a\cos\frac{1}{2}p_{x}a+2t_{1}\cos k_{y}a\cos \frac{1}{2}p_{y}a\\\nonumber&+2t_{2}\cos(k_{x}+k_{y})a\cos \frac{1}{2}(p_{x}+p_{y})a\\\nonumber&+2t_{\perp}\cos k_{z}d\cos \phi_{\perp},\\\nonumber
d_{x}({\mathbf k})&=t\cos\frac{1}{2}\left(k_x-k_y\right)a,\\\nonumber  
d_{y}({\mathbf k})&=-2t\sin\frac{1}{2}\left(k_x+k_y\right)a+t\sin\frac{1}{2}(k_{x}-k_{y})a,\\\nonumber   
d_{z}({\mathbf k})&=2t_{1}\sin k_{x}a\sin\frac{1}{2}p_{x}a+2t_{1}\sin k_{y}a\sin \frac{1}{2}p_{y}a\\\nonumber&+2t_{2}\sin(k_{x}+k_{y})a\sin \frac{1}{2}(p_{x}+p_{y})a\\&-2t_{\perp}\sin k_{z}d\sin \phi_{\perp}+\epsilon, \label{d}
\end{align}
where $a$ and $d$ are lattice constants in the $xy$-plane and along $\hat{z}$-axis. A $k_{z}$-dependent Zeeman term $m=\epsilon-2t_{\perp}\sin k_{z}d\sin \phi_{\perp}$ appears in the effective Zeeman fields along $\hat{z}$-direction. For fixed $m$, the Chern number
\begin{align}
C=\frac{1}{4\pi}\int dk_{x}dk_{y}\hat{d}({\bf k})\cdot (\partial_x \hat{d}({\bf k})\times \partial_x \hat{d}({\bf k}))
\end{align}
can be defined as the topological invariant of the system with fixed $k_z$ and $\hat{d}({\bf k})\equiv{\bf d}({\bf k})/|{\bf d}({\bf k})|$. 

\begin{figure}
\begin{center}
\includegraphics[width=3.3in]{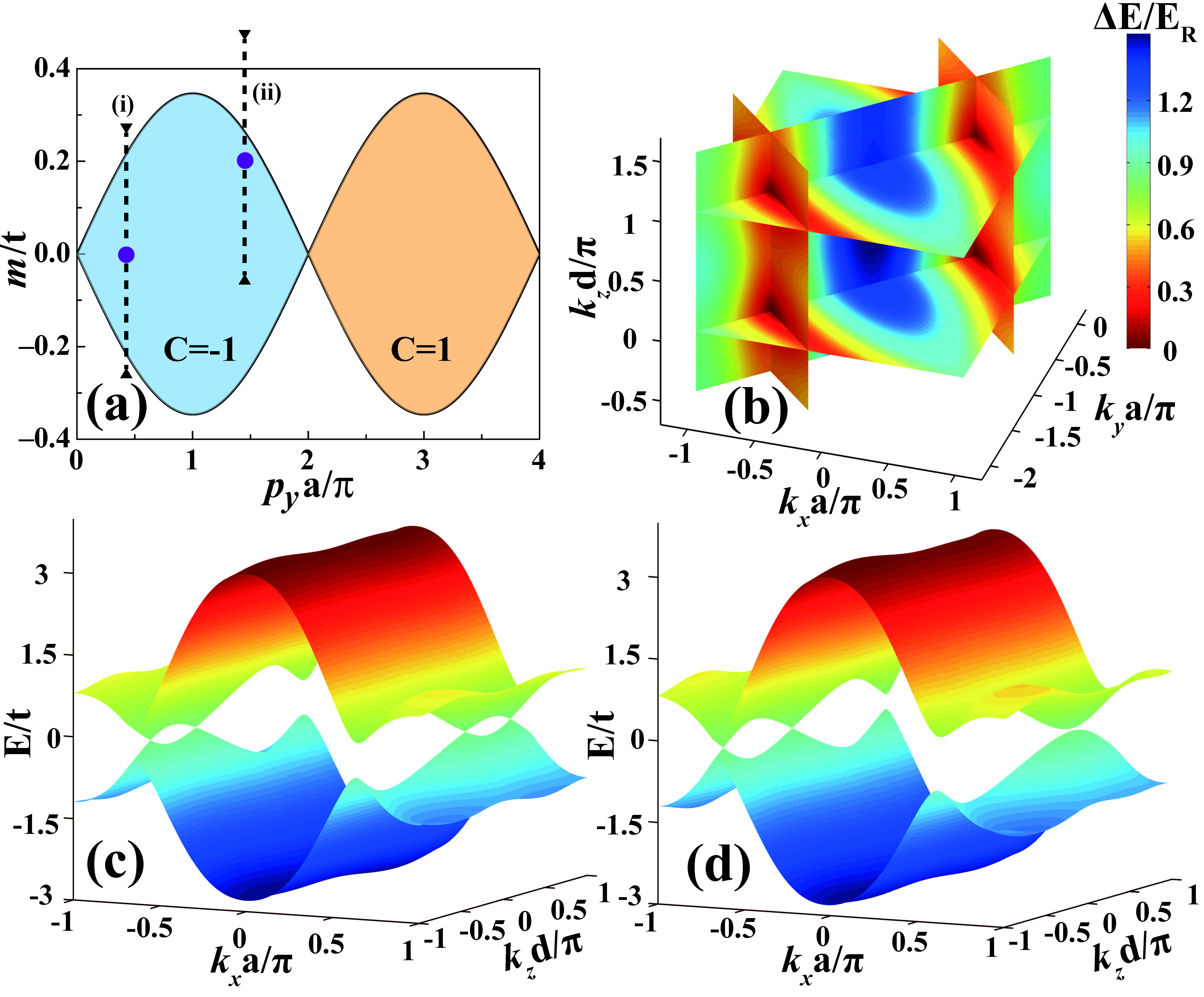}
\caption{(color online) (a) The Chern number $C$ of an isolated layer with $t_2=3t_1=0.3t$ as a function of the Zeeman term $m$ and recoil momentum $\Delta\mathbf{p}=\left(0,p_y,0\right)$. The purple dots denote $m(k_z=0)=\epsilon$, while the dashed lines correspond to the range of $m$ for general $k_{z}$ when interlayer coupling is turned on. (b) The energy difference between the upper and the lower bands, using parameters from a realistic optical lattice, is shown. It is evident that two pairs of Weyl points locate in the plane of $k_{y}=k_{x}-\pi$. (c) Energy spectrum in the $k_{y}=k_{x}-\pi$ plane with $\epsilon=0$ where two pairs of Weyl points emerge. (d) One pair of Weyl points emerges with $\epsilon=0.2t$. In (c) and (d) we take $t_1=0.1t$, $t_2=0.3t$, $t_{\perp}=0.26t$ and $\Delta\mathbf{p}=\left(0,\pi/a, \pi/d\right)$ for calculation.
}
\label{Fig2}
\end{center}
\end{figure}

The phase diagram of an effective 2D system with $\Delta\mathbf{p}=\left(0,p_y,0\right)$ as a function of fixed $m$ and $p_{y}$ is summarized in Fig.\ref{Fig2}a, where the colored region has non-zero Chern number. When the interlayer coupling $t_{\perp}$ is zero, $m=\epsilon$ with $p_ya \in \left(0, 2\pi\right)\cup\left(2\pi, 4\pi\right)$ makes $|d_z|>0$ for all $k_z$. As a result, the system is fully gapped in the bulk. 

As the interlayer coupling is increased to a finite value, $m$ as well as the bulk gap change as a function of $k_{z}$. From Eq.\ref{d}, it is clear that the Weyl points appear when $\left(k_xa,k_ya\right)=\left(-\frac{2}{3}\pi,-\frac{5}{3}\pi\right)$ or $\left(k_xa,k_ya\right)=\left(\frac{2}{3}\pi,-\frac{1}{3}\pi\right)$ and at a value of $k_{z}$ where $d_z(k_z)=0$ such that $|{\bf d}| =0$. For systems with specific $t_{\perp}$ and $\Delta\mathbf{p}$, the ranges of $m(k_z)$ as a function of $k_z$, where $k_z$ ranges from $-\pi/d$ to $\pi/d$, are schematically indicated as vertical dashed line (i) for $\epsilon=0$ and (ii) for $\epsilon=0.2t$ in Fig.\ref{Fig2}a. The number of Weyl points is determined by the number of crossings of the dashed lines and the topological phase boundary in Fig.\ref{Fig2}a. The crossing is controllable by tuning the interlayer coupling, the detuning and the momentum transfer $\Delta{\bf p}$ in the Raman process, so the Weyl semimetal phase with either one or two pairs of Weyl points becomes available. The energy spectrum for systems which contain one or two pairs of Weyl points are depicted in Fig.\ref{Fig2}c and d, respectively. The range of the Zeeman terms $m$ as a function of $k_z$ in Fig.\ref{Fig2}c and d correspond to the vertical lines (i) and (ii) in Fig.\ref{Fig2}a respectively.

\begin{figure}
\begin{center}
\includegraphics[width=3.3in]{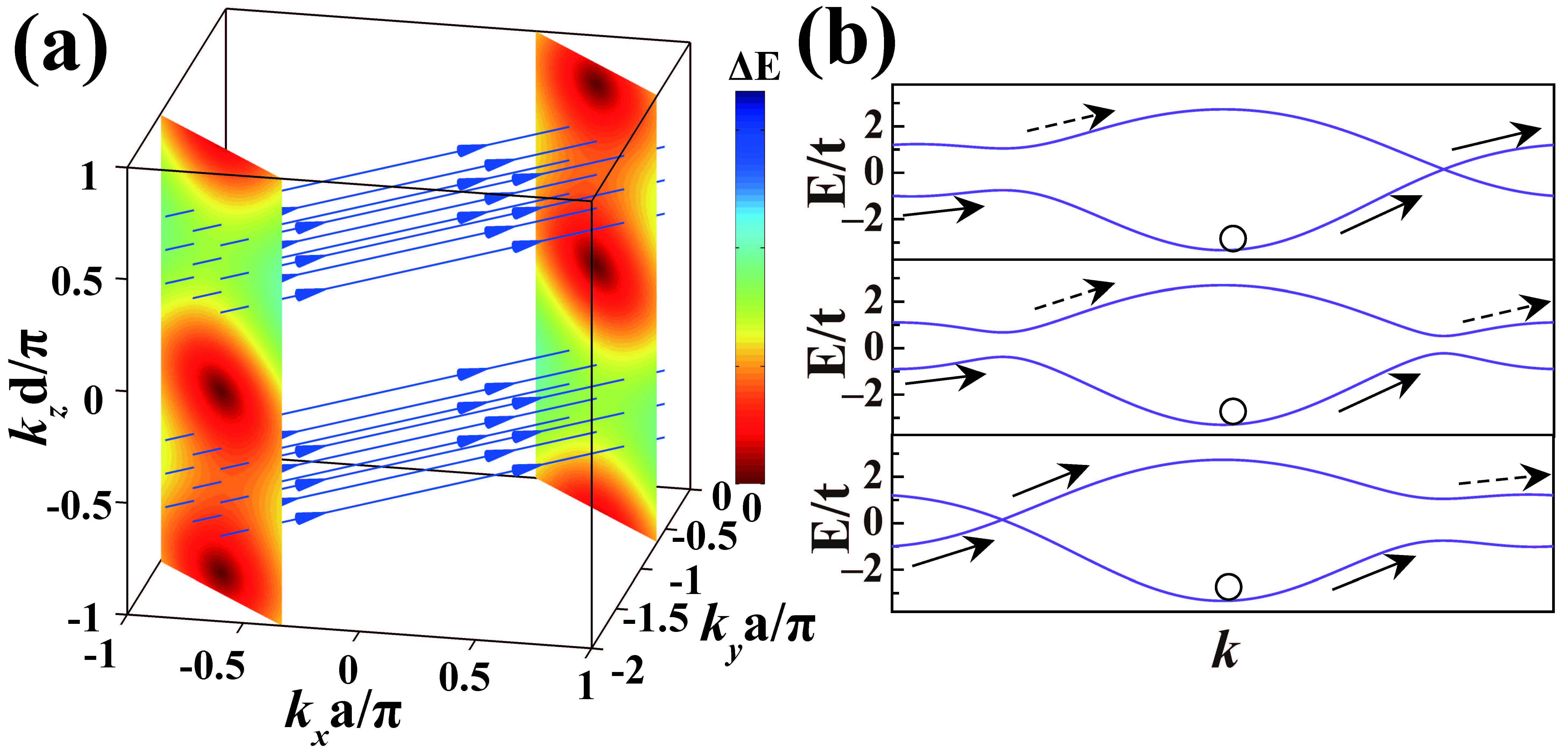}
\caption{(color online) (a) The two planes which contain the Weyl points are depicted. The color represents the energy difference between upper band and lower band and indicate the position of the Weyl points. Driven in the (110) direction, atoms at the bottom of the lower band will move along the arrows in the $\mathbf{k}$ space and penetrate the two planes. Atoms in a unit Bloch oscillation encounter two local band gap minima and undergo two Landau Zener tunneling events.  (b) Three typical cases of Landau-Zener transition. The path in the upper and bottom panel contains Weyl points with different locations, while that in the middle panel is fully gapped. The dashed arrow denotes that the Landau-Zener transition probability is small due to the existence of the energy gap.}
\label{Fig3}
\end{center}
\end{figure}

In the following, we consider a realistic optical potential with $\theta=\frac{2}{3}\pi$ and $V_{[\bar{X},X,Y]}=[7, 0.5, 2]E_{R}$, where $E_{R}\equiv q^2/2m$ is the recoil energy \cite{Tarruell,Uehlinger}. The hopping parameters of the optical lattices are calculated from the maximally localized Wannier function method (see Fig.\ref{FigS3} in the Appendix). In realistic band structures with the Raman laser strength $V_{0}=E_{R}$, two pairs of Weyl points emerge successively when the amplitude of $V_Z$ is increased. Using $t_{\perp}=0.09E_R, \Delta\mathbf{p}=\left(0, 2\pi/a, \pi/d\right), \epsilon=0$, we plot the energy gap between the lower and the upper bands in Fig.\ref{Fig2}b. Here the energy gap is plotted for the planes $k_z=0$, $k_z=\pi$, $k_y=k_x-\pi$ and $k_y=-k_x+k_{xi}+k_{yi}$ with $i=1,2$. The four crossing points $\left(k_{xi}, k_{yi}, 0\right)$ and $\left(k_{xi}, k_{yi}, \pi\right)$ between the orthogonal planes are the Weyl points. It is evident that the four band touching points represent the Weyl points. It is important to note that, using the realistic parameters, there is a wide range of experimentally accessible parameter space of $E_R$ and $V_z$ in which the Weyl semimetal phase can be realized.

\section{\bf Detection of Weyl semimetal phase}
Our proposed scheme for realizing Weyl semimetal consists of two steps: 1) create isolated 2D Chern insulators; and 2) couple the isolated layers with non-trivial hoppings. To confirm the emergence of the Weyl phase, we shall first detect the Chern insulator and then ascertain the closing of bulk gap when interlayer tunneling is increased. As we show below, both can be carried out with Bloch oscillations.

It is important to note that the nonzero Chern number comes from the integral of Berry curvature over the first Brillouin zone. Thus one way to reveal the topological order is to measure the Berry curvature, as was proposed theoretically in \cite{Price,Liu5}, and performed experimentally~\cite{Tarruell,Uehlinger}. In the Bloch oscillation induced by a constant force, atom cloud, in addition to the group velocity parallel to the force, would obtain an anomalous velocity that is transverse to the force due to nonzero Berry curvature \cite{Xiao}. After one full Bloch cycle, this Berry curvature generated anomalous velocity would slightly drive the atom cloud transversely. This transverse drift, which is measurable experimentally, can manifest the topological order of Chern insulators.

After the confirmation of 2D Chern insulator phase for the isolated layers, the interlayer coupling can be gradually enhanced through modulating the intensity of laser $\bar{Z}$ to close the bulk band gap \cite{Uehlinger}. Such band gap closing can be monitored by the fraction of atoms in the excited band after a cycle of Bloch oscillations, as was done for isolated 2D honeycomb lattice~\cite{Tarruell,Lim}. Below, we extend this technique to Weyl semimetal case and calculate the Landau-Zener probability. 
  
More specifically, we use the model described by $H(\mathbf k)$ in Eq.\ref{H} and consider an external linear potential applied along $(110)$ direction where there are two pairs of Weyl points, see Fig.\ref{Fig3}a. The total probability for inter-band transition will come from two independent Landau Zener events as schematically depicted in Fig.\ref{Fig3}b. Since the transition probability in each Landau-Zener event increases exponentially as the gap decreases, the emergence of Weyl points is accompanied by a dramatic increase in transferred fraction when bulk band gap closes. For noninteracting fermionic atoms, as was performed in recent experiment~\cite{Tarruell}, the Landau Zener probability must be averaged over the initial distribution of atoms~\cite{Lim}.

\begin{figure}
\begin{center}
\includegraphics[width=3.3in]{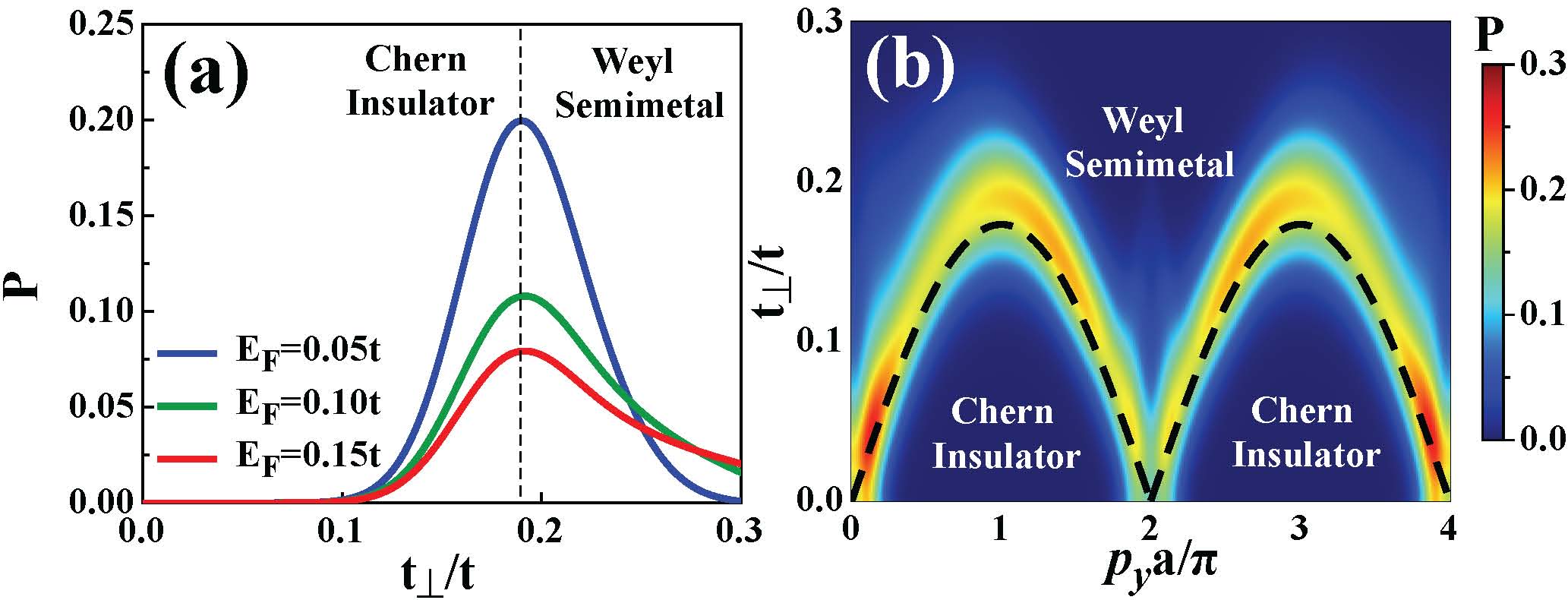}
\caption{(color online) (a) Landau-Zener transition probabilities as a function of $t_{\perp}$ for different Fermi energies. A transition peak signals the emergence of Weyl points at the critical value of $t_{\perp}$. The peak separates the fully gaped Chern insulator phase from the Weyl semimetal phase. (b) Landau-Zener transition probability as a function of both $t_{\perp}$ and $p$ with Fermi energy $E_{F}=0.05t$. In calculations, we take $F=0.01$, $t_1=0.1t$, $t_2=0.3t$, $\epsilon=0$ and $\Delta\mathbf{p}=\left(0, \pi/a, \pi/d\right)$.}
\label{Fig4}
\end{center}
\end{figure}

Let us denote the energy of the lowest band as $E({\bf k})$ and the energy difference between the lowest and the first excited bands close to band touching point as $\delta({\bf k})$. Then within local density approximation, the Landau-Zener transition probability can be written as
\begin{align}
P=\frac{\int_{E\left ( \mathbf{k},\mathbf{r} \right )\leq E_F}p({\bf k})d^3{\bf k}d^3{\bf r}}{\int_{E( \mathbf{k},\mathbf{r})\leq E_F}d^3{\bf k}d^3{\bf r}},
\end{align}
where $E_F$ is the Fermi energy and $E({\bf k},{\bf r})=E({\bf k})+V({\bf r})$ with $V({\bf r})$ the external harmonic confining potential. $p({\bf k})$ is the Landau-Zener probability for ${\bf k}$-state,
\begin{align}
p({\bf k})=\sum_{i\neq j}\exp\Big[-\frac{\pi\delta_i^2({\bf k})}{4v_iF}\Big]\Big(1-\exp\Big[-\frac{\pi\delta_j^2({\bf k})}{4v_jF}\Big]\Big),
\end{align}
where $i(j)$ labels the corresponding Landau-Zener events along $(110)$ direction and $v_{i(j)}$ is the velocity close to Weyl points. $F$ is the strength of the linear potential.  

The transition probability as a function of $t_{\perp}$, for systems with different $E_F$, is shown in Fig.\ref{Fig4}a. The momentum transfer due to the Raman field are $p_ya = \pi$ and $p_zd=\pi$ such that the system can be a Weyl semimetal when $t_{\perp}$ is larger than some critical values. Initially, atoms with energy lower than $E_F$ are occupied near the $\vec{k}=\left(0,0,\pm\pi/2d\right)$. For small $t_{\perp}$, the system is fully gapped and the Landau-Zener transition probability is exponentially small. At certain critical $t_{\perp}$, Weyl points emerge on the line along the $(110)$ direction. Due to the gap closing, the inter-band transition probability increases dramatically as shown in Fig.\ref{Fig4}a. However, by further increasing $t_{\perp}$, the Weyl points move away from $\left|k_zd\right|=\pi/2$ plane. As a result, the atoms driven in the $(110)$ direction will not be able to access the Weyl points and the transition amplitude decreases. Therefore, there is a peak in the transition probability as a function of $t_{\perp}$ and the peak location indicates the critical value of $t_{\perp}$ at which the Weyl points emerge. In other words, the transition peak separates the fully gapped Chern insulator phase from the nodal Weyl semimetal phase.

The density plot of the transition probabilities as a function of $t_{\perp}$ and $p_y$ for Fermi energy $E_F=0.05t$ is shown in Fig.\ref{Fig4}b. The Landau-Zener transition peak locates around $t_{\perp}=\frac{\sqrt{3}}{2}\left|\sin\frac{1}{2}p_ya\left(t_1-t_2\right)\right|$, which is denoted as the dashed line in Fig.\ref{Fig4}b. As in Fig.\ref{Fig4}a, the peak separates the Chern insulator phase from the Weyl semimetal phase. Similar results can be obtained when realistic parameters are used.

\begin{figure}
\begin{center}
\includegraphics[width=3.3in]{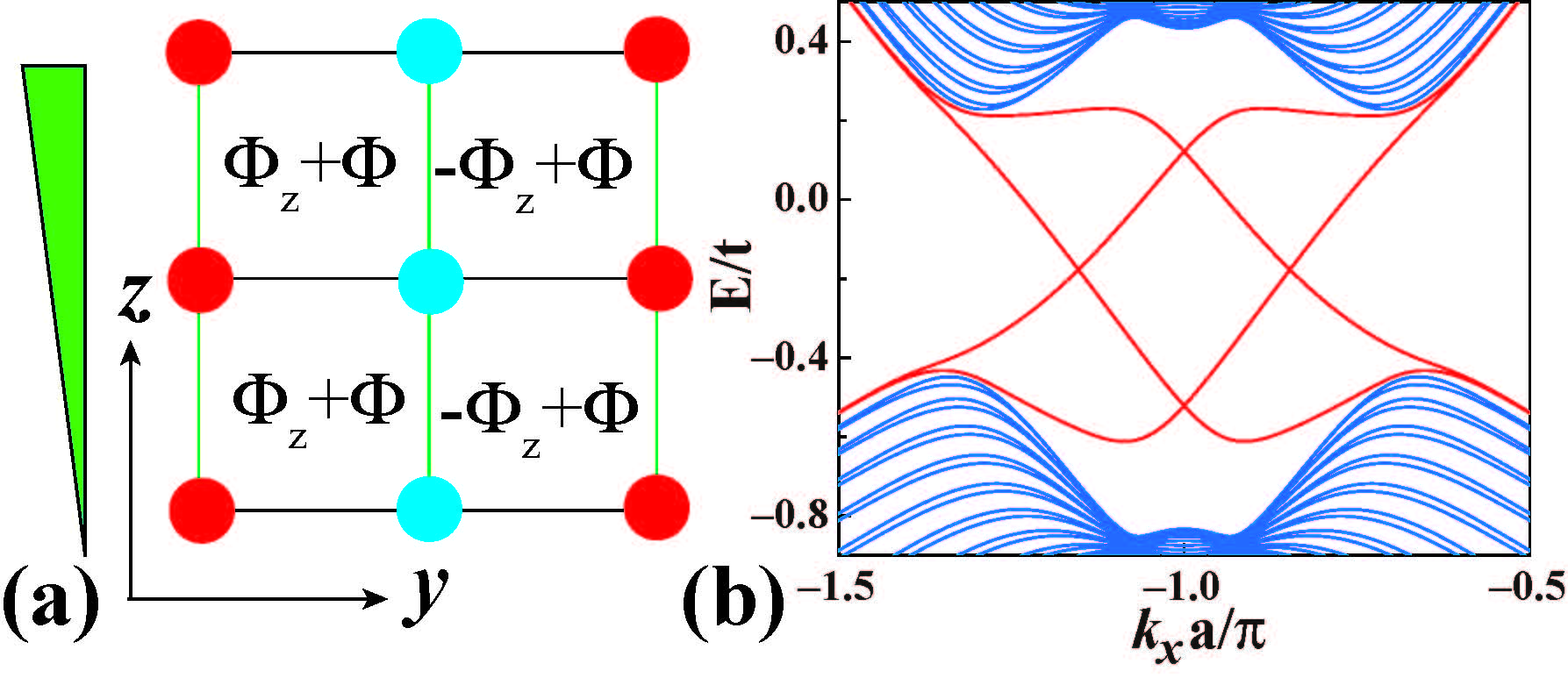}
\caption{(color online) (a) Schematic illustration of uniform magnetic flux $\Phi$ generated by a magnetic field gradient in the $z$-direction and the Raman laser assisted tunnelings. This flux $\Phi$ coexist with $\Phi_z$ which is responsible for the emergence of the Weyl points. (b) The energy spectrum of the Weyl semimetal in the presence of $\Phi$. The in-gap gapless modes (denoted by the red lines) connect the gapped Weyl points with opposite chirality. The in-gap gapless modes denotes the energy spectrum formed by the zeroth Landau levels. }
\label{Fig5}
\end{center}
\end{figure}

\section{\bf Landau Levels and Chiral Anomaly}
It is important to note that the Weyl semimetal phase discussed above can be used to realize the chiral anomaly. Considering a low-energy Weyl Hamiltonian without any velocity anisotropy for simplicity, in the presence of an external magnetic field $\mathbf{B}=\left(B_x, 0, 0\right)$, we can choose the Landau gauge $\mathbf{A}=\left(0, 0, eBy\right)$ to obtain $H_{\text{W}}=\chi v_F\left[\delta k_x\sigma_x-i\partial_y\sigma_y+\left(\delta k_z+eBy\right)\sigma_z\right]$, where $\chi=\pm1$ represents the chirality of the Weyl point. Along the direction parallel to the magnetic field, the infinite Landau bands form gapped spectrum $E_n=\hbar v_F \def\sgn{\mathop{\rm sgn}\nolimits} \sgn \left(n\right)\sqrt{2\left| n\right| eB/\hbar+\delta k_x^2}$, with $n=\pm1, \pm2, ...$ and also gapless chiral modes $E_0=\chi v_F\delta k_x$. The gapless chiral modes induced by the zeroth Landau levels have opposite group velocity for the states at Weyl points of opposite chirality. When a constant force (gradient scalar field) is applied parallel to the magnetic field, states near one Weyl point can be driven to the other Weyl point of opposite chirality through the gapless chiral modes. In this case the particle number is not conserved in a single Weyl point but the total particle number over the two Weyl points remains a constant. In the background of the magnetic field and gradient scalar field, this violation of particle number in a single Weyl point of specific chirality is refered as chiral anomaly~\cite{Nielsen}. In the following, we will show that a uniform synthetic magnetic field can be introduced in our Weyl semimetal phase to form the chiral Landau bands and further simulate the chiral anomaly.

In our AA stacked honeycomb optical lattice, when a linear magnetic field gradient is induced along the $z$-direction and together with the laser assisted tunneling, an extra phase $\Phi=\Delta{\bf p}\cdot{\bf r}_{i,j,l}=i\phi_x+j\phi_y+l\phi_z$ is added to the $z$ direction hopping $t^{a\left(b\right)}_{\perp}$~\cite{Aidelsburger1}. In our choice of Raman lasers, $\left(\phi_x,\phi_y,\phi_z\right)=\left(0,\frac{\pi}{2},\pi\right)$, the extra phase $\Phi$ results in a uniform synthetic magnetic field along the $x$ direction, as it is schematic shown in Fig.\ref{Fig5}a.

In the presence of this uniform magnetic field along the $x$-direction, energy spectrum formed by Landau levels are generated. The zeroth Landau levels result in gapless chiral modes along the $k_x$-direction at the Weyl points. The chiralities of the gapless chiral modes are determined by the chiralities of the Weyl points~\cite{potter}. The energy levels in the presence of the synthetic magnetic flux as a function of $k_x$ are depicted in Fig.\ref{Fig5}b. The in-gap modes represent the energy of the zeroth Landau levels. 

With a constant force applied parallel to the magnetic field along the $x$-direction, atoms can be adiabatically pumped from one Weyl point to another Weyl point with opposite chirality. This is the manifestation of chiral anomaly. The chiral anomaly induced atom number imbalance between different Weyl points is expected to cause asymmetric quasi-momentum distribution in the time-of-flight measurements~\cite{Struck, Spielman, Fallani}. Furthermore the recent advancement in cold atom quantum transport measurement~\cite{Chien, Brantut, Krinner} can enable the study of chiral anomaly induced negative magnetoresistance~\cite{Chen} in atomic systems.

\section{\bf Conclusion}
In conclusion, we introduced an experimentally feasible way for realising Weyl semimetals by applying Raman lasers to coupled multilayer honeycomb optical lattices. We suggest that the measurements of Landau-Zener tunneling probabilities can be used to detect the Weyl semimetal phase and the possible exploration of chiral anomaly is discussed. Our scheme will lay the foundation for the realisation of even more exotic nodal topological phases such as Weyl superconductors~\cite{Balents,Vincent} when attractive interactions between atoms are introduced.

\section*{\bf Acknowledgement}
We thank the support of HKRGC through HKUST3/CRF/13G. W.-Y.H. and K.T.L are further supported by GRF Grants 602813, 605512, 16303014. S.Z.Z is further supported by HKU709313P. 

{\em Note Added}: At the finishing stage of this work, we noted that a scheme to realize Weyl semimetal phase by stacking 2D Harper systems in cubic lattices was proposed~\cite{Buljan}. 

\appendix
\setcounter{figure}{0}
\renewcommand{\thefigure}{A\arabic{figure}}

\section{\bf TIGHT BINDING MODEL} \label{App-A}
\begin{figure}
\begin{center}
\includegraphics[width=3.3in]{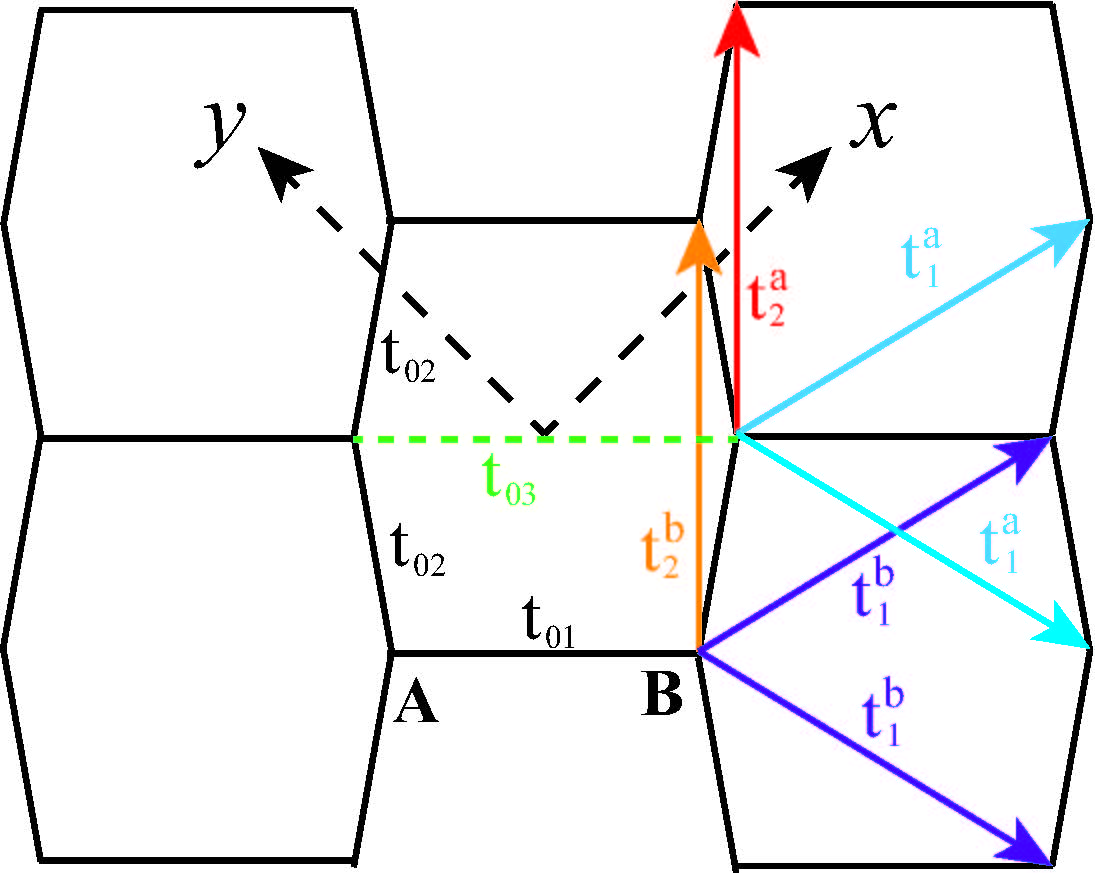}
\caption{The hopping terms considered in the honeycomb lattice (brickwall lattice).}
\label{FigS1}
\end{center}
\end{figure}

Coupled by Raman field, the periodic drive system involving the resonant modulation can be described by a time-independent effective Hamiltonian~\cite{Jaksch,Cooper}. With hopping terms indicated in Fig.\ref{FigS1}, the Hamiltonian reads
\begin{align}\nonumber
H&=\sum_{m,n,l}t_1^a\left(a_{m-1,n,l}^{\dagger}a_{m,n,l}+a_{m,n-1,l}^{\dagger}a_{m,n,l} \right )\\\nonumber&+t_2^aa_{m-1,n+1,l}^{\dagger}a_{m,n,l}+t_{\perp}a_{m,n,l-1}^{\dagger}a_{m,n,l}\\\nonumber&+t_1^b\left(b_{m-1,n,l}^{\dagger}b_{m,n,l}+b_{m,n-1,l}^{\dagger}b_{m,n,l} \right )\\\nonumber&+t_2^bb_{m-1,n+1,l}^{\dagger}b_{m,n,l}+t_{\perp}b_{m,n,l-1}^{\dagger}b_{m,n,l}\\\nonumber&+\left[2V_0\sin\left(\Delta\mathbf{p}\cdot \mathbf{r}_{m,n,l}^a-\delta\omega t \right ) -\frac{\Delta}{2}\right ]a_{m,n,l}^{\dagger}a_{m,n,l}\\\nonumber&+\left[2V_0\sin\left(\Delta\mathbf{p}\cdot \mathbf{r}_{m,n,l}^b-\delta\omega t \right ) +\frac{\Delta}{2}\right ]b_{m,n,l}^{\dagger}b_{m,n,l}\\\nonumber&+t_{02}\left(a_{m,n,l}^{\dagger}b_{m-1,n,l}+a_{m,n,l}^{\dagger}b_{m,n-1,l} \right )\\&+t_{01}a_{m,n,l}^{\dagger}b_{m,n,l}++h.c.,
\end{align}
where $r_{m,n,l}^a=m\mathbf{v}_1+n\mathbf{v}_2+l_\mathbf{d}$ and $r_{m,n,l}^b=m\mathbf{v}_1+\mathbf{v}_2+l\mathbf{d}+\mathbf{R}_{b}-\mathbf{R}_{a}$. Considering the gauge transformation
\begin{align}\nonumber
U&=\sum_{m,n,l}\exp\left(\frac{1}{2}i\delta\omega t\right)a_{m,n,k}^{\dagger}a_{m,n,k}\\&+\exp\left(-\frac{1}{2}i\delta\omega t\right)b_{m,n,k}^{\dagger}b_{m,n,k},
\end{align}
we have
\begin{align}\nonumber
H'&=U^{\dagger}HU-i\hbar U^{\dagger}\frac{\mathrm{d} U}{\mathrm{d}t}\\&=H_0+H_1e^{i\delta\omega t}+H_{-1}e^{-i\delta\omega t},
\end{align}
with
\begin{align}\nonumber
H_0&=\sum_{m,n,l}t_1^a\left(a_{m-1,n,l}^{\dagger}a_{m,n,l}+a_{m,n-1,l}^{\dagger}a_{m,n,l} \right )\\\nonumber&+t_2^aa_{m-1,n+1,l}^{\dagger}a_{m,n,l}+t_{\perp}a_{m,n,l-1}^{\dagger}a_{m,n,l}\\\nonumber&+t_1^b\left(b_{m-1,n,l}^{\dagger}b_{m,n,l}+b_{m,n-1,l}^{\dagger}b_{m,n,l} \right )\\\nonumber&+t_2^bb_{m-1,n+1,l}^{\dagger}b_{m,n,l}+t_{\perp}b_{m,n,l-1}^{\dagger}b_{m,n,l}\\
&+\epsilon\left(a_{m,n,l}^{\dagger}a_{m,n,l}-b_{m,n,l}^{\dagger}b_{m,n,l}\right)+h.c.,
\end{align}
\begin{align}\nonumber
H_1&=\sum_{m,n,l}t_{01}b_{m,n,l}^{\dagger}a_{m,n,l}\\\nonumber&+t_{02}\left(b_{m-1,n,k}^{\dagger}a_{m,n,k}+b_{m,n-1,k}^{\dagger}a_{m,n,k}\right)\\\nonumber&+V_0e^{-i\left(\Delta\mathbf{p}\cdot\mathbf{r}_{m,n,l}^a-\frac{\pi}{2}\right)}a_{m,n,l}^{\dagger}a_{m,n,l}\\
&+V_0e^{-i\left(\Delta\mathbf{p}\cdot\mathbf{r}_{m,n,l}^b-\frac{\pi}{2}\right)}b_{m,n,l}^{\dagger}b_{m,n,l},
\end{align}
and
\begin{align}\nonumber
H_{-1}&=\sum_{m,n,l}t_{01}a_{m,n,l}^{\dagger}b_{m,n,l}\\\nonumber&+t_{02}\left(a_{m-1,n,k}^{\dagger}b_{m,n,k}+a_{m,n-1,k}^{\dagger}b_{m,n,k}\right)\\\nonumber&+V_0e^{i\left(\Delta\mathbf{p}\cdot\mathbf{r}_{m,n,l}^a-\frac{\pi}{2}\right)}a_{m,n,l}^{\dagger}a_{m,n,l}\\
&+V_0e^{i\left(\Delta\mathbf{p}\cdot\mathbf{r}_{m,n,l}^b-\frac{\pi}{2}\right)}b_{m,n,l}^{\dagger}b_{m,n,l}.
\end{align}
Here we consider the effective Hamiltonian to the first order
\begin{align}
H_{\text{eff}}=H_0+\frac{1}{\hbar\omega}\left[H_1,H_{-1}\right]+O\left(\frac{1}{\omega^2}\right),
\end{align}
then
\begin{align}
H_{\text{eff}}^{\left(0\right)}&=H_0,\\\nonumber
H_{\text{eff}}^{\left(1\right)}&=\frac{1}{\hbar\omega}\sum_{m,n,l}\left(t_{01}^2+2t_{02}^2\right)b_{m,n,l}^{\dagger}b_{m,n,l}\\\nonumber&+t_{01}t_{02}\left(b_{m,n,l}^{\dagger}b_{m-1,n,l}+b_{m,n,l}^{\dagger}b_{m,n-1,l}\right)\\\nonumber&+t_{02}^2b_{m,n,l}^{\dagger}b_{m+1,n-1,l}-\left(t_{01}^2+2t_{02}^2\right)a_{m,n,l}^{\dagger}a_{m,n,l}\\\nonumber&-t_{01}t_{02}\left(a_{m,n,l}^{\dagger}a_{m-1,n,l}+a_{m,n,l}^{\dagger}a_{m,n-1,l}\right)\\\nonumber&-t_{02}^2a_{m,n,l}^{\dagger}a_{m+1,n-1,l}\\\nonumber
&+2t_{02}V^{m,n}_{m-1,n}e^{-i\frac{1}{2}\Delta\mathbf{p}\cdot\left(\mathbf{r}_{m,n,l}^a+\mathbf{r}_{m-1,n,l}^b\right)}a_{m,n,l}^{\dagger}b_{m-1,n,l}\\\nonumber
&+2t_{02}V^{m,n}_{m-1,n}e^{i\frac{1}{2}\Delta\mathbf{p}\cdot\left(\mathbf{r}_{m,n,l}^a+\mathbf{r}_{m-1,n,l}^b\right)}b_{m-1,n,l}^{\dagger}a_{m,n,l}\\\nonumber
&+2t_{02}V^{m,n}_{m,n-1}e^{-i\frac{1}{2}\Delta\mathbf{p}\cdot\left(\mathbf{r}_{m,n,l}^a+\mathbf{r}_{m,n-1,l}^b\right)}a_{m,n,l}^{\dagger}b_{m,n-1,l}\\\nonumber
&+2t_{02}V^{m,n}_{m,n-1}e^{i\frac{1}{2}\Delta\mathbf{p}\cdot\left(\mathbf{r}_{m,n,l}^a+\mathbf{r}_{m,n-1,l}^b\right)}b_{m,n-1,l}^{\dagger}a_{m,n,l}\\\nonumber
&+2t_{01}V^{m,n}_{m,n}e^{-i\frac{1}{2}\Delta\mathbf{p}\cdot\left(\mathbf{r}_{m,n,l}^a+\mathbf{r}_{m,n,l}^b\right)}a_{m,n,l}^{\dagger}b_{m,n,l}\\
&+2t_{01}V^{m,n}_{m,n}e^{i\frac{1}{2}\Delta\mathbf{p}\cdot\left(\mathbf{r}_{m,n,l}^a+\mathbf{r}_{m,n,l}^b\right)}b_{m,n,l}^{\dagger}a_{m,n,l},
\end{align}
\begin{figure}
\begin{center}
\includegraphics[width=3.3in]{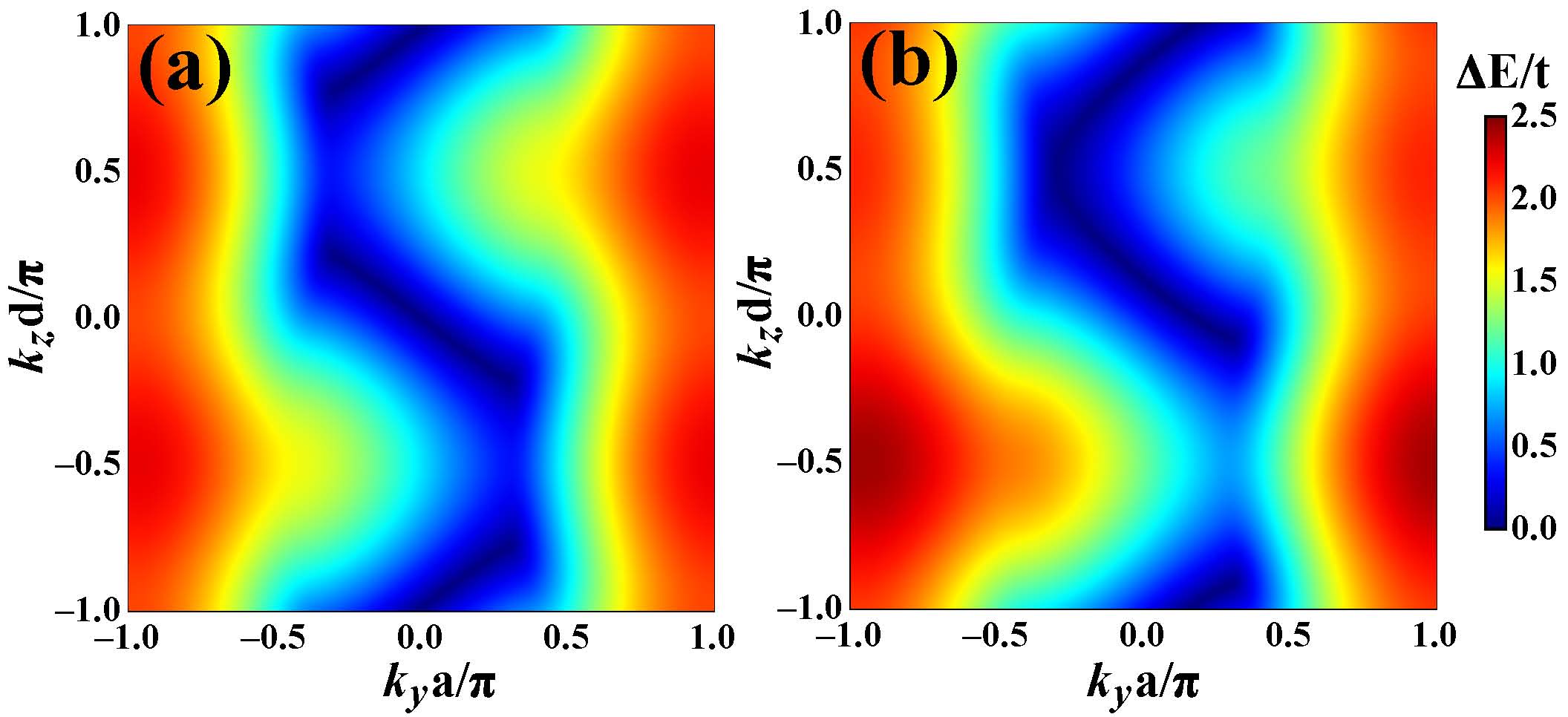}
\caption{The Fermi arc on the (100) surfaces of Weyl semimetals. (a) using the parameters corresponding to Fig.2c in the main text. (b) using the parameters corresponding to Fig.2d in the main text. }
\label{FigS2}
\end{center}
\end{figure}
where $V^{m,n}_{m',n'}=V_0\sin\frac{1}{2}\Delta\mathbf{p}\cdot\left(\mathbf{r}_{m,n,l}^a-\mathbf{r}_{m',n',l}^b\right)$.It is clear that the Raman lasers will introduce the Peierls phase $\exp\left[i\Delta\mathbf{p}\cdot\left(\mathbf{r}+\mathbf{r}'\right)/2\right]$. For the ideal model in the maintext, we assume $t_1^a=t_1^b=t_1$, $t_2^a=t_2^b=t_2$, and ignore the difference in AA and BB sublattice hopping in the effective Hamiltonian. For $\Delta\mathbf{p}=\left(0,p_y,p_z\right)$, $\left|\sin\Delta\mathbf{p}\cdot\left(\mathbf{r}_{i,l}-\mathbf{r}_{j,l}\right)\right|$ has the same value for the nearest neighbor hopping, so the total Hamiltonian is simplified as
\begin{align}\nonumber
H=&\sum_{\left \langle i,j \right \rangle,l}\left[t_{ij}^0e^{i\frac{1}{2}\Delta\mathbf{p}\cdot\left(\mathbf{r}_{i,l}^a+\mathbf{r}_{j,l}^b\right)}b_{i,l}^{\dagger}a_{j,l}+t_{ij}^aa_{i,l}^{\dagger}a_{j,l}+t_{ij}^bb_{i,l}^{\dagger}b_{j,l}\right]\\\nonumber
&+\epsilon\sum_{i,l}\left(a_{i,l}^{\dagger}a_{i,l}-b_{i,l}^{\dagger}b_{i,l}\right)\\&+\sum_{i,l}\left(t_{\perp}^aa_{i,l}^{\dagger}a_{i,l+1}+t_{\perp}^bb_{i,l}^{\dagger}b_{i,l+1}\right)+h.c..
\end{align}
The spatial dependence of the Peierls phases is eliminated through the unitary transformation as follows
\begin{figure}
\begin{center}
\includegraphics[width=3.3in]{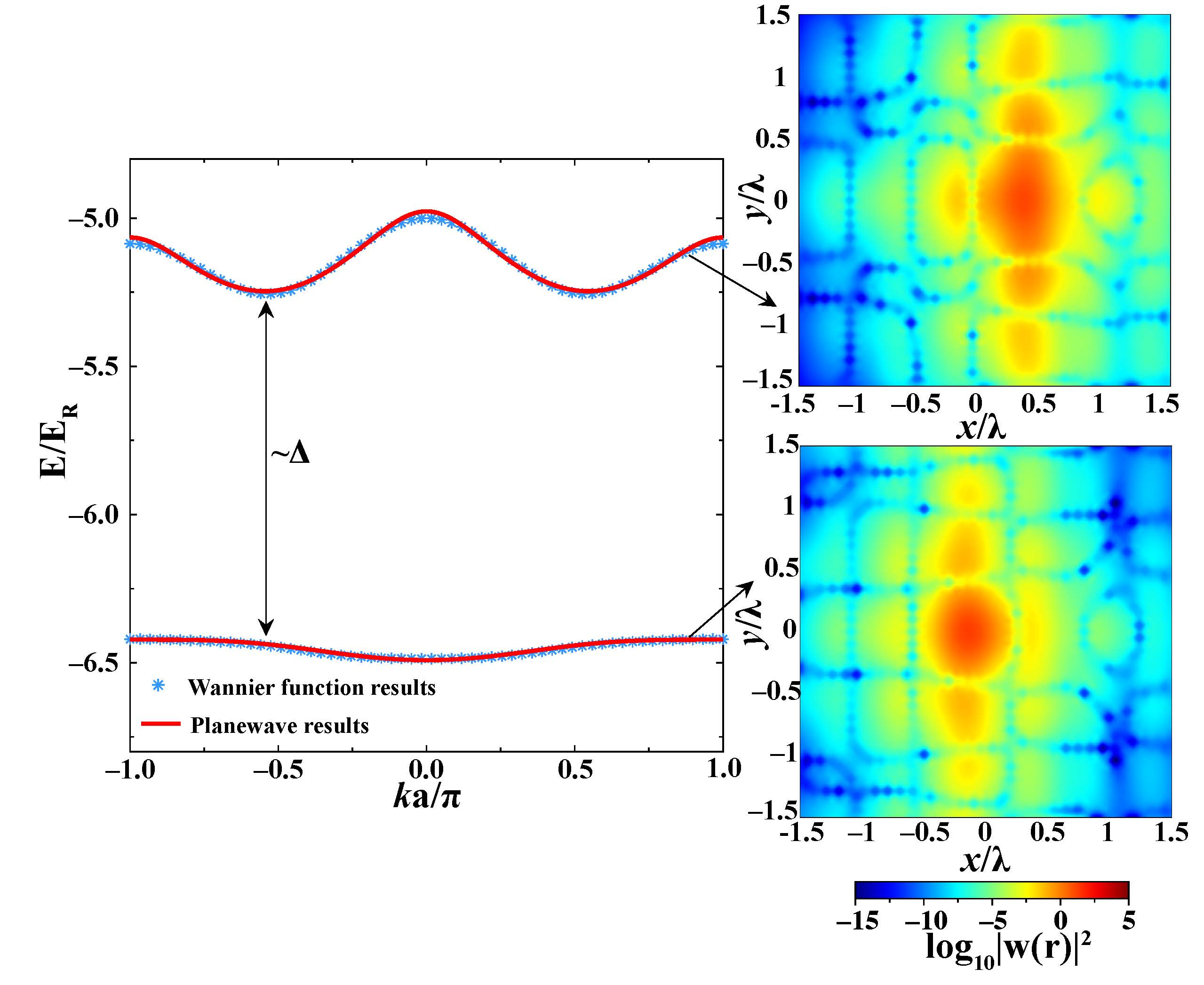}
\caption{The band structure and its corresponding maximally localized Wannier function for the lattice potential with $V_{\bar{X},X,Y}=\left[7,0.5,2\right]E_R$ and $\theta=\frac{2}{3}\pi$. The hopping terms obtained from maximally localized Wannier function makes the energy spectrum agree well with that from plane wave results. $\mathbf{k}$ is along $k_x=k_y$. $\lambda=\frac{2\pi}{q}=\sqrt{2}a$ is the wavelength of the in-plane laser.}
\label{FigS3}
\end{center}
\end{figure}
\section{WEYL POINTS IN REALISTIC OPTICAL POTENTIAL} \label{App-B}
\begin{align}\nonumber
\tilde{a}_{i,l}&=a_{i,l}\exp\left(i\frac{1}{2}\Delta\mathbf{p}\cdot\mathbf{r}_{i,l}^a\right)\\
\tilde{b}_{j,l}&=b_{j,l}\exp\left(-i\frac{1}{2}\Delta\mathbf{p}\cdot\mathbf{r}_{j,l}^b\right).
\end{align}
After the gauge transformation, the new Hamiltonian takes the form
\begin{align}\nonumber
H=&\sum_{ \langle i,j \rangle, l }\left ( t_{ij}^{0}\tilde{b}_{i,l}^{\dagger}\tilde{a}_{j,l}+ { t_{ij}^{a}e^{i\phi _{ij}}\tilde{a}_{i,l}^{\dagger}\tilde{a}_{j,l}} + { t_{ij}^{b}e^{-i\phi _{ij}}\tilde{b}_{i,l}^{\dagger}\tilde{b}_{j,l}}  \right ) \\\nonumber
&+\epsilon \sum_{i,l} \left ({ \tilde{a}_{i,l}^{\dagger}\tilde{a}_{i,l}}\right. \left.{-\tilde{b}_{i,l}^{\dagger}\tilde{b}_{i,l}} \right ) +\sum_{i,l} \left ({ {t_{\perp }^{a}e^{i\phi _{\perp}}\tilde{a}_{i,l}^{\dagger}\tilde{a}_{i,l+1}}}\right.\\
&\left.{+ t_{\perp }^{b}e^{-i\phi _{\perp}}\tilde{b}_{i,l}^{\dagger}\tilde{b}_{i,l+1}}\right )+ h.c..
\end{align}
Further labeling $A\left(B\right)$ sublattices as corresponding to a pseudo-spin $\mathbf{\sigma}$, then the Hamiltonian is transformed in k space as follows
\begin{align}
H\left(\mathbf{k}\right)=\begin{pmatrix}
h_{11}&h_{12} \\ h_{21}
& h_{22}
\end{pmatrix}=d_0\left(\mathbf{k}\right)I+\mathbf{d}\left(\mathbf{k}\right)\cdot\bm{\sigma}.
\end{align}
Weyl points will emerge when the conditions $d_x\left(\mathbf{k}\right)=d_y\left(\mathbf{k}\right)=d_z\left(\mathbf{k}\right)=0$ are satisfied. For the case $\left|t^0_{ij}\right|=t$, $d_x\left(\mathbf{k}\right)=d_y\left(\mathbf{k}\right)=0$ occurs when $\left(k_xa,k_ya\right)=\left(-\frac{2}{3}\pi,-\frac{5}{3}\pi\right)$ or $\left(k_xa,k_ya\right)=\left(\frac{2}{3}\pi,-\frac{1}{3}\pi\right)$. Thus the critical condition for the emergence of Weyl points can be determined by judging the existence of real roots for $d_z\left(-\frac{5}{3}\pi,-\frac{2}{3}\pi,k_z\right)=0$ or $d_z\left(-\frac{1}{3}\pi,\frac{2}{3}\pi,k_z\right)=0$ with given $\epsilon$ and $\phi_{\perp}$.Physically the existence of Weyl points corresponds to a transition from topological nontrivial state to topological trivial state when atoms move along $k_z$ direction. The Fermi arc evolution is shown in Fig.\ref{FigS2}.

Realistic optical lattice potential with $V_{\left[\bar{X},X,Y\right]}=\left[7,0.5,2\right]E_R$ and $\theta=\frac{2}{3}\pi$ is considered to calculate the band structure for isolated layer and its corresponding Maximally localized Wannier function~\cite{Walters}. The Wannier function $W\left(x,y,z\right)=w\left(x,y\right)w\left(z\right)$ is controlled by in-plane and out of plane optical potential respectively, and $w\left(x,y\right)$ is shown in Fig.\ref{FigS3} with intralayer hopping parameters $\left[t_{01},t_{02},t_{03},t_1^a,t_2^a,t_1^b,t_2^b\right]=10^{-2}\left[-8.29,-1.65,0,0.45,1.86,-0.21,3.22\right]E_R$ and the onsite potential $\Delta=-1.2E_R$. In the resonant case $\Delta=\delta\omega$, we can select the wavelength for the z direction standing wave to have $p_zd=\pi$ and make $V_0=E_R$. Such optical potential can generate Weyl points as is shown in Fig.\ref{Fig2}b.

\end{document}